\documentclass{sig-alternate-05-2015}
\usepackage{amsmath}
\usepackage{mathptmx}
\usepackage{balance} 
\usepackage{graphicx}
\usepackage{color}
\usepackage{xspace}
\usepackage[T1]{fontenc}
\usepackage{textcomp}
\usepackage{times}
\usepackage[captionskip=0pt]{subfig}
\DeclareRobustCommand{\insertpicture}[2]{
	\begin{center}\includegraphics[width=#2 \textwidth]{#1}
	\end{center}} 
\DeclareRobustCommand{\etal}{\textit{et al.}\xspace}
\DeclareRobustCommand{\bp}{\emph{backward projection}\xspace}
\DeclareRobustCommand{\Bp}{\emph{Backward projection}\xspace}
\DeclareRobustCommand{\fp}{\emph{forward projection}\xspace}
\DeclareRobustCommand{\Fp}{\emph{Forward projection}\xspace}
\DeclareRobustCommand{\pl}{\emph{prolines}\xspace}
\DeclareRobustCommand{\Pl}{\emph{Prolines}\xspace}
\DeclareRobustCommand{\surl}[1]{\emph{\urlstyle{same}\url{#1}}\xspace}
\definecolor{orange}{rgb}{1, 0.49, 0.05}
\definecolor{gray}{rgb}{0.49, 0.49, 0.49}

\DeclareRobustCommand{\cagatay}{\c{C}a\u{g}atay Demiralp\xspace}
\newcommand*{\modern}{\fontfamily{cmss}\selectfont}

\begin{document}\sloppy
\setcopyright{acmcopyright}


%
\conferenceinfo{KDD 2016 Workshop on Interactive Data Exploration and Analytics (IDEA'16)}{August 14th, 2016, San Francisco, CA, USA.}
\CopyrightYear{2016}

\title{Clustrophile: A Tool for Visual Clustering Analysis}

\numberofauthors{1} 
\author{
\alignauthor \cagatay \\
       \affaddr{IBM Research}\\
       \email{cagatay.demiralp@us.ibm.com
       }
}
\maketitle
\begin{abstract}
 While clustering is one of the most popular methods for data mining, analysts 
 lack adequate tools for quick, iterative clustering analysis, which
 is essential for hypothesis generation and data reasoning. We introduce
 Clustrophile, an interactive tool for iteratively computing discrete and
 continuous data clusters, rapidly exploring different choices of clustering
 parameters, and reasoning about clustering instances in relation to data
 dimensions. Clustrophile combines three basic visualizations -- a table of raw
 datasets, a scatter plot of planar projections, and a matrix diagram (heatmap)
 of discrete clusterings -- through interaction and intermediate visual encoding.
 Clustrophile also contributes two spatial interaction techniques,
 \emph{forward projection} and \emph{backward projection}, and a visualization
 method, \emph{prolines}, for reasoning about two-dimensional projections
 obtained through dimensionality reductions. 
 \end{abstract}

\keywords{
  Clustering, 
  projection, 
  dimensionality reduction, 
  visual analysis,
  experiment, 
  Tukey,
  out-of-sample extension, 
  forward projection, 
  backward projection, 
  prolines, 
  sampling, 
  scalable visualization,
  interactive analytics.}


\section{Introduction}

Clustering is a basic method in data mining. By automatically dividing data
into subsets based on similarity, clustering algorithms provide a simple yet
powerful means to explore structures and variations in data. 
What makes clustering attractive is its unsupervised (automated) nature, which
reduces the analysis time. Nonetheless, analysts need to make several
decisions on a clustering analysis that determine what constitutes a cluster,
including which clustering algorithm and similarity measure to use, which
samples and features (dimensions) to include, and what granularity (e.g.,
number of clusters) to seek.  Therefore, quickly exploring the effects of
alternative decisions is important in both reasoning about the data and making
these choices.

Although standard tools such as R or Matlab are extensive and computationally
powerful, they are not designed to support such interactive iterative analysis.
It is often cumbersome, if not impossible,  to run what-if scenarios with these
tools. In response, we introduce Clustrophile, an interactive visual analysis
tool, to help analysts to perform iterative clustering analysis.  Clustrophile
couples three basic visualizations, a dynamic table listing of raw datasets, a
scatter plot of planar projections, and a matrix diagram (heatmap) of discrete
clusterings, using interaction and intermediate visual encoding. We consider
dimensionality reduction as a form of continuous clustering that complements
the discrete nature of standard clustering techniques.  We also contribute 
two spatial interaction techniques, \fp  and \bp, and a visualization method, 
\pl, for reasoning about two-dimensional projections computed using dimensionality 
reductions.  

\begin{figure*}[t]
 \centering
 \insertpicture{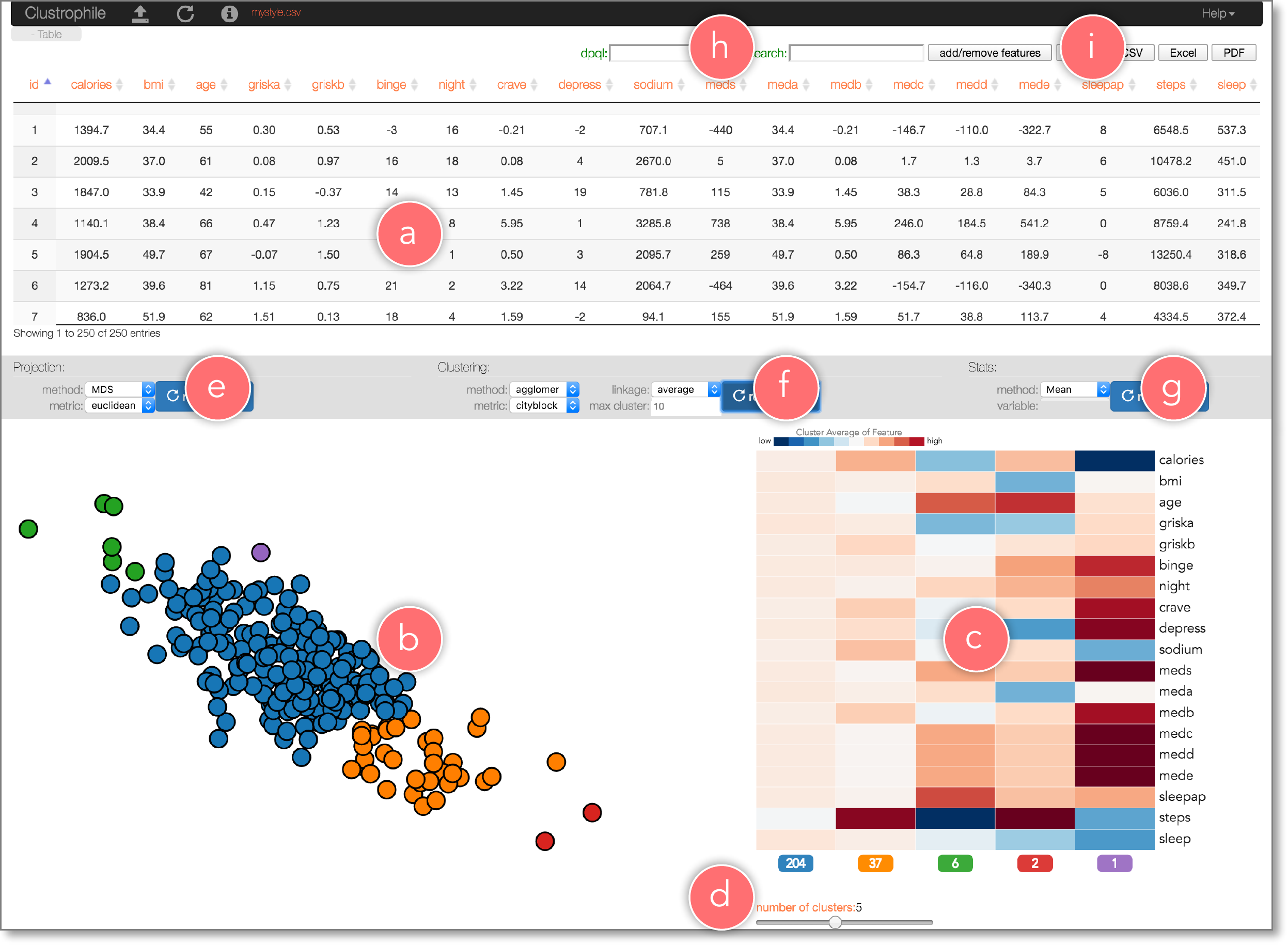}{0.9}
 \caption{
     Clustrophile is an interactive visual analysis tool for computing data
     clusters and iteratively exploring and reasoning about clustering
     instances in relation to data subsets and dimensions through what-if
     scenarios.  To this end, Clustrophile combines three basic visualizations,
     a) a table of raw datasets, b) a scatter plot of planar projections, and
     c) a matrix diagram (heatmap) of discrete clusterings, using interaction
     and intermediate visual encoding.  Clustrophile enables users to
     interactively d) change the number of clusters, quickly explore several e)
     projection and f) clustering algorithms and parameters, run g) statistical
     analysis, including hypothesis testing,  and  dynamically filter h) the
     observations and i) features to which visual analysis is applied.  
     \label{fig:interface}}
\end{figure*}

\section{Related Work} 

Clustrophile builds on earlier work on interactive systems supporting visual
clustering analysis. The projection interaction and visualization techniques 
in Clustrophile are related to prior efforts in user experience with 
scatter-plot visualizations of dimensionality reductions.

\subsection{Visualizing Clusterings}  

Prior research applies visualization for improving user understanding of
clustering results across domains. Using coordinated visualizations with
drill-down/up capabilities is a typical approach in earlier interactive
tools. The Hierarchical Clustering Explorer ~\cite{Jinwook_Seo_2002} is an
early and comprehensive example of interactive visualization tools for
exploring clusterings. It supports the exploration of hierarchical
clusterings of gene expression datasets through dendrograms (hierarchical
clustering trees) stacked up with heatmap visualizations.   

Earlier work also proposes tools that make it possible to incorporate user
feedback into clustering formation.  Matchmaker~\cite{Lex_2010} builds on
techniques from~\cite{Jinwook_Seo_2002} with the ability to modify clusterings
by grouping data dimensions.  ClusterSculptor~\cite{Nam_2007} and Cluster
Sculptor ~\cite{Bruneau_2015}, two different tools, enable users to supervise
clustering processes in various clustering methods.  Schreck \etal
~\cite{Schreck_2009} propose using user feedback to bootstrap the similarity
evaluation in data space (trajectories, in this case) and then apply the
clustering algorithm. 

Prior work has also introduced techniques for comparing clustering
results of different datasets or different
algorithms~\cite{Cao_2011,Lyi_2015,Pilhofer_2012,Jinwook_Seo_2002}.
DICON~\cite{Cao_2011} encodes statistical properties of clustering instances as
icons and embeds them in the plane based on similarity using multidimensional
scaling.  Pilhofer \etal \cite{Pilhofer_2012} propose a method for reordering
categorical variables to align with each other and thus augment the visual
comparison of clusterings. The recent tool XCluSim~\cite{Lyi_2015} supports
comparison of several clustering results of gene expression datasets using an
approach similar to that of the Hierarchical Clustering Explorer.   

Clustrophile is similar to earlier work in coordinating basic and auxiliary
visualizations to explore clusterings. Clustrophile focuses on supporting
iterative, interactive exploration of data with the ability to explore
multiple choices of algorithmic parameters along with hypothesis testing
through visualizations and interactions as well as formal statistical
methods. Finally, Clustrophile is domain-agnostic and is intended to be a
general tool for data scientists.  

\begin{figure*}[t]
  \centering
\includegraphics[height=0.29\textwidth]{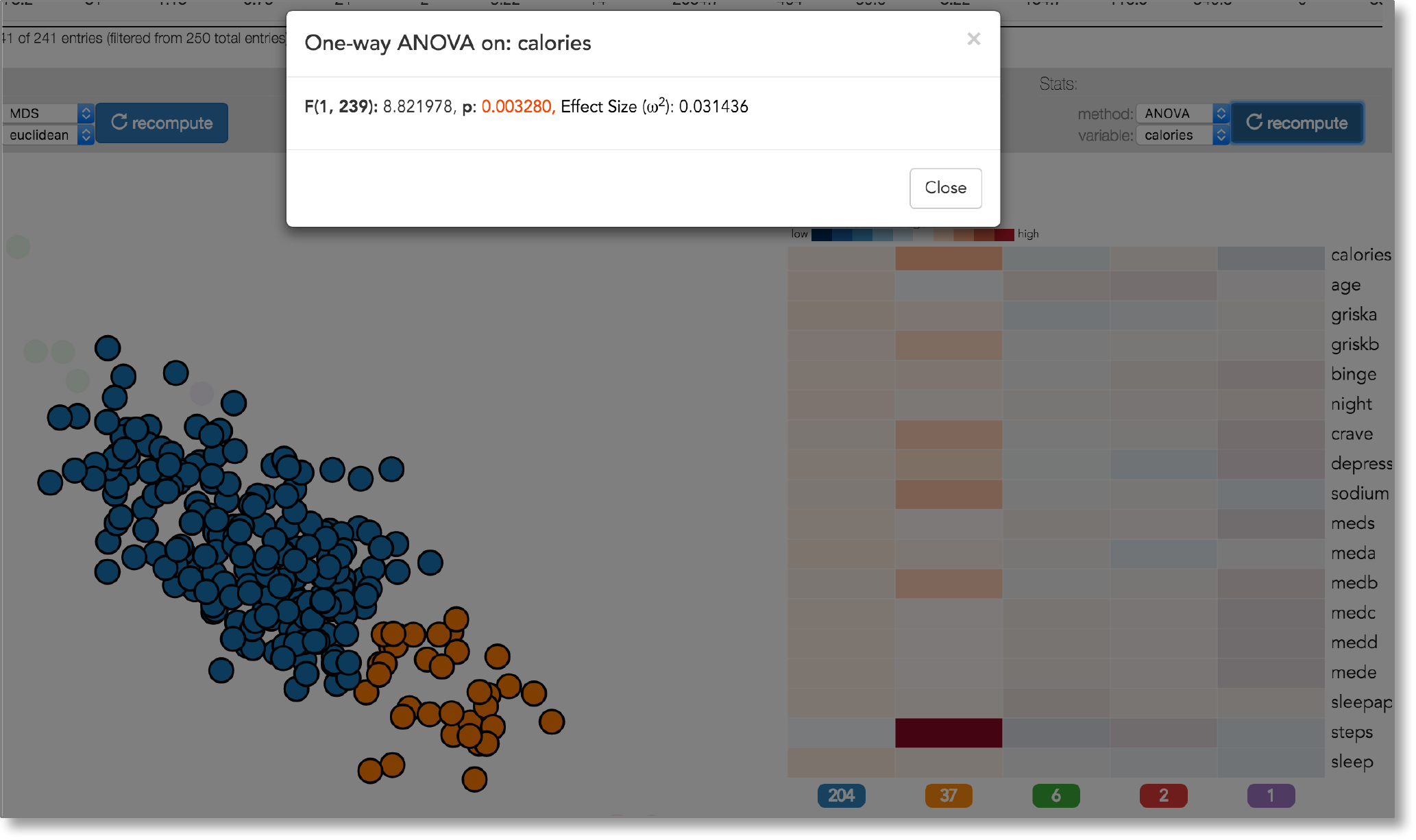}
\includegraphics[height=0.29\textwidth]{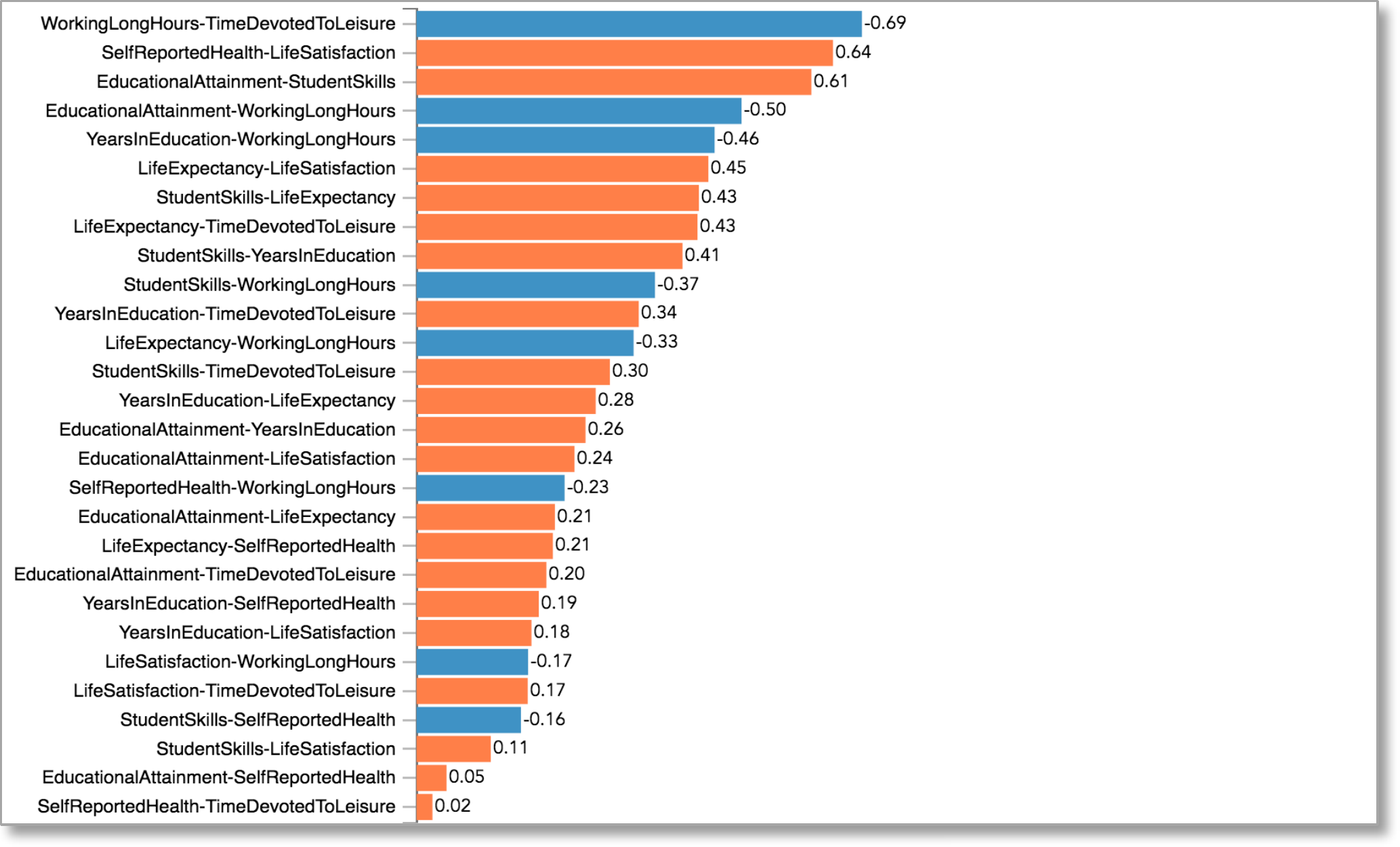}
\caption{ (Left) ANOVA test on {\modern calories} between two selected clusters 
in a life style dataset. (Right) Correlation coefficients for all pairs of features in a development 
indicators dataset~\cite{oecd} for OECD member states.
Correlation values are sorted based on their absolute value. The sign of
correlation is encoded by color.  
\label{fig:stat}} 
\end{figure*}

\subsection{Making Sense with and of Dimensionality Reductions}  

Dimensionality reduction is a common method for analyzing and visualizing
high-dimensional datasets across domains. Researchers in statistics and
psychology pioneered the use of techniques that project multivariate data onto
low-dimensional manifolds for visual analysis
(e.g.,~\cite{Asimov_1985,Friedman_1974,PRIM9_1974,kruskal:mds64,shepard:mds62,torgerson:pm52}).
PRIM-9 (Picturing, Rotation, Isolation, and Masking --- in up to 9 dimensions)
~\cite{PRIM9_1974} is an early visualization system supporting exploratory data
analyis through projections. PRIM-9 enables the user to interactively rotate
the multivariate data  while continuously viewing a two dimensional projection
of the data. Motivated by the user behavior in the PRIM-9 system, Friedman and
Tukey ~\cite{Friedman_1974} first propose a measure, the projection index, for
quantifying the ``usefulness'' of a given projection plane (or line) and, then,
an optimization method, the projection pursuit, to find the most useful
projection direction (i.e., one that has the highest projection index value).
The proposed index considers the projections that result in large spread with
high local density to be useful (e.g., highly separated clusters). In an axiomatic 
approach that complements the
projection pursuit, Asimov introduces the grand tour, a method for viewing
multidimensional data via orthogonal projections onto a sequence of
two-dimensional planes~\cite{Asimov_1985}. Asimov considers a set of criteria
such as density, continuity, and uniformity to select a sequence of projection
planes from all possible projection planes and provides specific methods to
devise such sequences. Note that the space of all possible two-dimensional
planes through the origin is a Grassmannian manifold.  Asimov's grand tours can
be seen as geodesic curves with desired properties in this manifold. 
 
Despite their wide use (and overuse), interpreting and reasoning about
dimensionality reductions can often be difficult.  Earlier work focuses on
better conveying projection (reduction) errors, integrating user feedback
into the projection process and evaluating the effectiveness of various
dimensionality reductions. Low-dimensional projections are generally lossy
representations of the data relations.  Therefore, it is useful to convey
both overall and per-point dimensionality reduction errors to users when
desired.  Earlier research proposes techniques for visualizing projection
errors using Voronoi diagrams~\cite{Aupetit_2007,Lespinats_2010} and
``correcting'' them within a neighborhood of the probed
point~\cite{Chuang_2012,Stahnke_2016}.  Stahnke \etal~\cite{Stahnke_2016}
suggest a set of interactive methods for interpreting the meaning and quality
of projections visualized as scatter plots. The methods make it possible to
see approximation errors, reason about positioning of elements, compare them
to each other, and visualize the extrapolated density of individual
dimensions in the projection space.  

In certain cases, expert users have prior knowledge of how the projections
should look. To enable user input to guide dimensionality reduction, earlier
research has proposed several techniques
\cite{Buja_2008,Endert_2012,Gleicher_2013,
Jeong_2009,Johansson_2009,Williams_2004}. Enabling users to adjust the
projection positions or the weights of data dimensions and distances is a
common approach in earlier research for incorporating user feedback  to
projection computations. For example, X/GGvis~\cite{Buja_2008} supports changing 
the weights of dissimilarities input to the MDS stress function along the with 
the coordinates (configuration) of the embedded  points to guide the projection 
process.  Similarly, iPCA~\cite{Jeong_2009} enables  users to interactively modify 
the weights of data dimensions in computing projections.  Endert
\etal \cite{Endert_2011} apply similar ideas to an additional set of
dimensionality-reduction methods while incorporating user feedback through
spatial interactions. The spatial interactions, \fp and \bp, that we introduce
here are developed for dynamically reasoning about dimensionality-reduction
methods and the underlying data, not for incorporating user feedback.  

Prior research also evaluates dimensionality-reduction 
techniques~\cite{Brehmer_2014,Lewis_2012} as well as visualization methods
for representing dimensionally-reduced data ~\cite{Sedlmair_2013}.  Sedlmair
\etal find that two-dimensional scatter plots outperform scatter-plot
matrices and three-dimensional scatter plots in the task of separating
clusters~\cite{Sedlmair_2013}. Lewis \etal \cite{Lewis_2012} report that
experts are consistent in evaluating the quality of dimensionality reductions
obtained by different methods, but novices are highly inconsistent in such
evaluations.  A later study finds, however, that experts with limited
experience in dimensionality reduction also lack clear understanding of
dimensionality-reduction results~\cite{Brehmer_2014}.  

\Fp, \bp and \pl are new techniques and complement earlier work in improving
interactive reasoning with dimensionality reductions, particularly in 
order to facilitate dynamically asking and answering hypothetical questions 
about both the underlying data and the dimensionality reduction.   

\section{The Design of Clustrophile}  

We developed Clustrophile for data scientists, using their regular feedback at
each stage of the development process. We discuss below the design of
Clustrophile, stressing the rationale behind our choices, basic visualizations
and interactions.     

\subsection{Design Criteria}
In our collaboration with data scientists, we identified four high-level
criteria to consider in designing Clustrophile. 

\textbf{Show Variation Within Clusters}     
Clustering is useful for grouping data points based on similarity, enabling
users to discover salient structures in data while reducing the cognitive load.
However, differences among data points within clusters are lost.  Clustrophile
has coordinated views---Table, Projection, and Clustering---that facilitate
exploration of differences among data points at different levels of
granularity. The projection view holds a scatter-plot visualization of the data
reduced to two dimensions through dimensional reduction, thus providing a
continuous spatial view of similarities among high-dimensional data points.  

\textbf{Allow Quick Iteration over Parameters}
In clustering analysis, users typically need to make several decisions,
including which clustering method and distance (dissimilarity) measure to use,
how many clusters to create, which features and data subsets to consider, and 
the like. After an initial clustering, users would like to be able to iterate on and
refine these decisions. Clustrophile enables users to interactively update and
apply clustering and projection algorithms and parameters at any point in
their analysis.

\textbf{Facilitate Reasoning about Clustering Instances}
Users often would like to know what features (dimensions) of the data points
are important in determining a given clustering instance or how different
choices of features or distance measures might affect the clustering.
Clustrophile allows users to add/remove features interactively and to change
distance measures used in clustering and projections. 

\textbf{Promote Multiscale Exploration}
The ability to interactively drill down into data is crucial for exploration
and effective use of visual encoding variables, particularly in two-dimensional
space.  Clustrophile supports dynamic filtering of data across the views.  In
addition, Clustrophile makes possible the application of clustering and
projection methods to filtered subsets of data, providing a semantic zoom-in
and zoom-out capability.    

\begin{figure} 
  \includegraphics[width=0.49\textwidth]{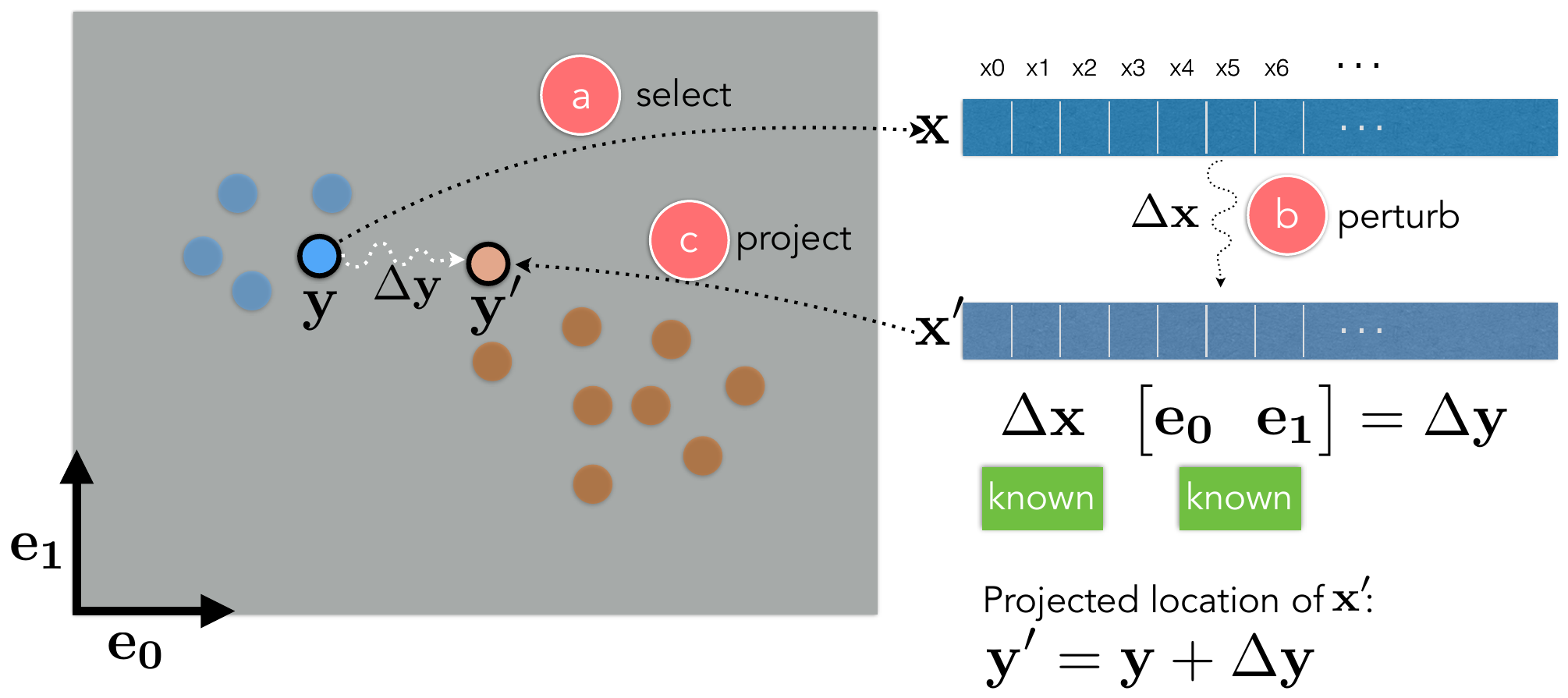} \caption{
    \Fp enables users to a) select any data point $\mathbf{x}$ in the 
    projection, b) interactively change the feature (dimension) values 
    of the point and c) observe how that changes the current projected 
    location $\mathbf{y}$ of the point. For PCA, the positional 
    change vector $\Delta \mathbf{y}$ can be derived directly by projecting 
    the data change vector $\Delta \mathbf{x}$ onto the first two principal 
    components, $\mathbf{e_0}$ and $\mathbf{e_1}$. \label{fig:fp}} 
\end{figure}
\subsection{Views}     
Clustrophile has five coordinated views: Table, Projection, Clustering, 
Statistics and Playground.   

\textbf{Table} The Table view (Figure~\ref{fig:interface}a) contains a
dynamic table visualization of data. Tables in Clustrophile can be searched,
filtered, sorted, and exported as needed (Figure~\ref{fig:interface}h,i). Upon
loading, data first appears as a table listing in this view, giving users a
direct and familiar way to access the records. Clustrophile supports input
files in the Comma Separated Values (CSV) format. Clustrophile also enables
exporting the current table in CSV, Portable Document Format (PDF), or Excel
file formats. Alternatively, users can simply the current table to the
clipboard to paste in other applications.  

\textbf{Clustering} The Clustering view (Figure~\ref{fig:interface}c) 
contains a heatmap (matrix diagram) visualization of the current clustering. 
The columns of the heatmap
corresponds to the number of clusters and are ordered from left to right based
on size  (i.e., the first column represents the largest cluster in the current
clustering). The rows of the heatmap represent the features, and the color of each
cell encodes the normalized average feature value for clusters.  
Clustrophile supports dynamic computation of clusterings using the kmeans and
agglomerative clustering algorithms with several choices of similarity measures
and, in the case of agglomerative clustering, linkage options (Figure~\ref{fig:interface}f).  
The choices can
be changed easily and clustering can be recomputed using the model panel above
the clustering view.  Similarly, users can dynamically change the number 
of clusters by using a sliding bar (Figure~\ref{fig:interface}d). 

\textbf{Projection} Clustering algorithms divide data into discrete
groups based on similarity, but different degrees of variation within and
between groups are suppressed. Clustrophile provides two-dimensional
projections obtained using dimensionality reduction that complement the
discrete clusterings. The Projection view
(Figure~\ref{fig:interface}b) contains a scatter-plot visualization of the
current data reduced to two dimensions by using one of six
dimensionality-reduction methods: Principal Component Analysis (PCA), Classical
Multidimensional Scaling (CMDS), non-metric Multidimensional Scaling (MDS),
Isomap, Locally Linear Embedding (LLE), and t-distributed Stochastic Neighbor
Embedding (t-SNE)~\cite{Maaten_2009}. As with clustering, users can select
among several similarity measures with which to run the projection algorithms
(Figure~\ref{fig:interface}e).  Each circle in the scatter plot 
represents a data point and their color encodes their cluster membership 
in the currently active clustering method. 

\textbf{Statistics} This view displays the results of 
the most recent statistical computation. Currently, Clustrophile provides 
standard point statistics along with a hypothesis-testing functionality 
using ANOVA and pairwise correlation computations between 
features (Figure~\ref{fig:stat}). 

\textbf{Playground} Clustrophile enables the exploration of two- 
dimensional projections of the data through forward and backward projections. 
In the Playground view, users can create a copy of an existing data point 
and interactively modify its feature values to see how its projected  
position changes. Conversely, users can change the projected position and 
see what feature values satisfy this change. 

\begin{figure}[t]
\centering
\includegraphics[width=0.45\textwidth]{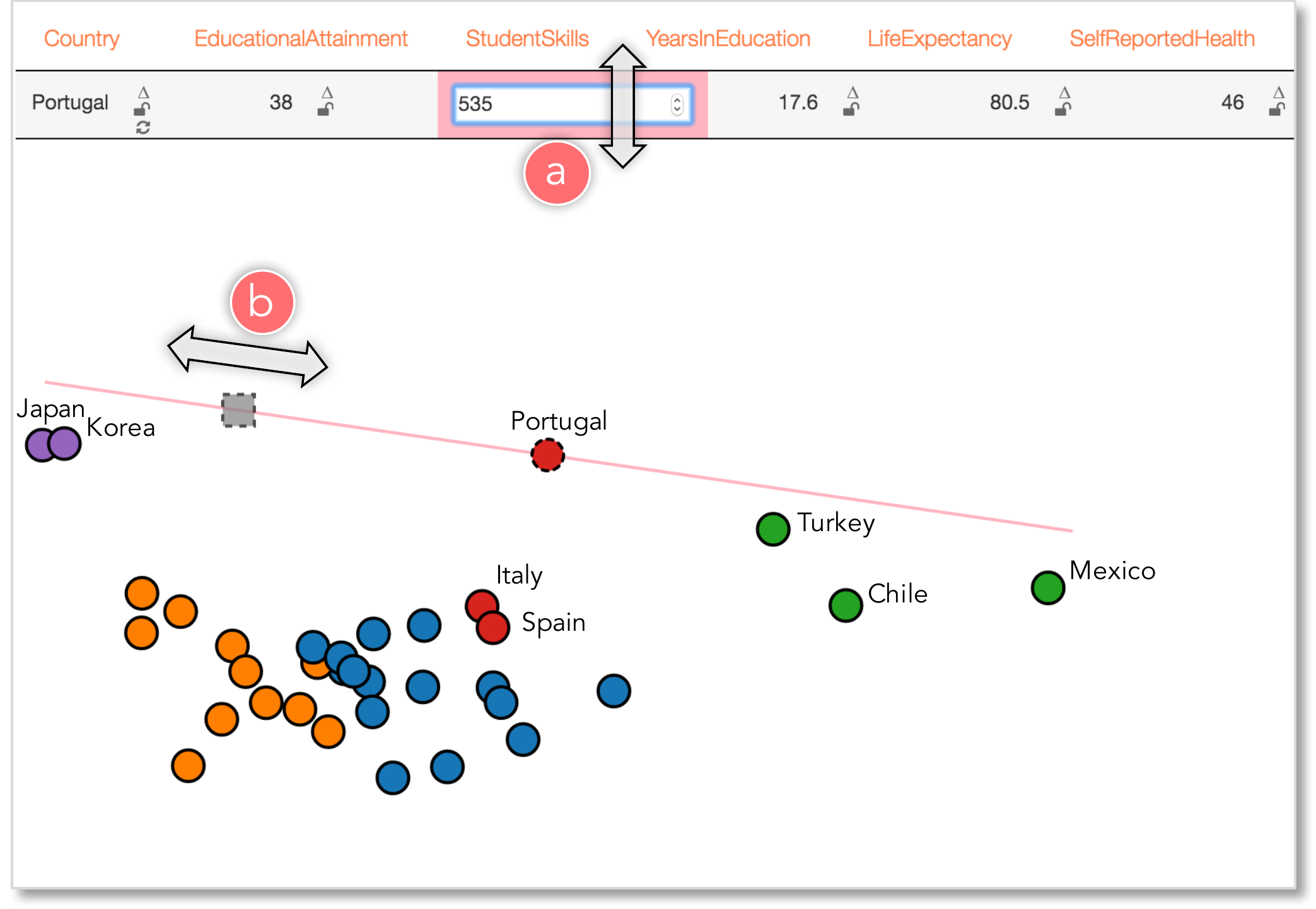}
\caption{\Fp in action. \Fp enables user to explore if and how much 
  {\modern StudentSkills} explains the difference between Portugal and 
  Korea in a projection of OECD member countries based on a set of 
  development indices. The user a) dynamically changes the value of 
  the {\modern StudenSkills} dimension for Portugal and b) observes 
  the dynamically updated projection. In this case, user discovers that 
  {\modern StudentSkills} is the most important feature explaining 
  the difference between Portugal and Korea.\label{fig:fpinaction}} 
\end{figure}

\subsection{Interactions} 

\textbf{Brushing and Linking}. 
We use brushing \& linking to select data across and coordinate the views 
of Clustrophile.  This is the main mechanism that lets users 
observe the effects of one operation across the views.

\textbf{Dynamic filtering}  
In addition to brushing, Clustrophile provides two basic mechanisms for
dynamically filtering data (Figure~\ref{fig:interface}h). First, its search 
functionality lets users filter the data using arbitrary keyword search on feature names and values.  Second, users can also
filter the table using expressions in a mini-language. For example, 
typing {\modern age > 40 \& weight<180} dynamical selects data points across 
views where the fields {\modern age} and {\modern weight} satisfy the 
entered constraint.

\textbf{Adding and Removing Features}
Understanding the relevance of data dimensions or features to the analysis is
an important yet challenging goal in data analysis. Clustrophile enables users
to add and remove features (dimensions) and explore the resulting changes in
clustering and projection results (Figure~\ref{fig:interface}i).

\subsection{Interacting with Dimensionality Reductions} 

Dimensionality reduction is the process of reducing the number of 
dimensions in a high-dimensional dataset in a way that maximally preserves
inter-datapoint relations of some form  as measured in the original
high-dimensional space. As with clustering, most dimensionality-reduction
techniques are unsupervised and learn salient structures explaining the data.
Unlike clustering, however, dimensionality-reduction methods discover
continuous representations of these structures. 

Despite its ubiquitous use, dimensionality reduction can be difficult to 
interpret, particularly in relation to original data dimensions. \emph{What do the
axes mean?} is probably users' most frequent question when looking at scatter
plots in which points (nodes) correspond to dimensionally-reduced data.
Clustrophile integrates  \fp, \bp, and \pl to facilitate direct, dynamic
examination of dimensionality reductions represented as scatter plots.  

There are many dimensionality-reduction
methods~\cite{Maaten_2009} and developing effective and scalable 
dimensionality-reduction algorithms is an active research area.   
Here we focus on principal component analysis (PCA), one of
the most frequently used dimensionality-reduction techniques; note that the
discussion here applies as well to other linear dimensionality-reduction 
methods. PCA computes (learns) a linear orthogonal transformation (high-dimensional rotation) 
of the empirically centered data into a new coordinate frame in which the axes represent maximal
variability. The orthogonal axes of the new coordinate frame are called
principal components. To reduce the number of dimensions to two, for example,
we project the centered data matrix, rows of which correspond to data
samples and columns to features (dimensions), onto the first two principal
components, $\mathbf{e_0}$ and $\mathbf{e_1}$. 
Details of PCA along with its many formulations and interpretations can be
found in standard textbooks on machine learning or data mining (e.g.,
\cite{Bishop_2006,Hastie_2005}).  

\begin{figure}[t]
\centering
\includegraphics[width=0.5\textwidth]{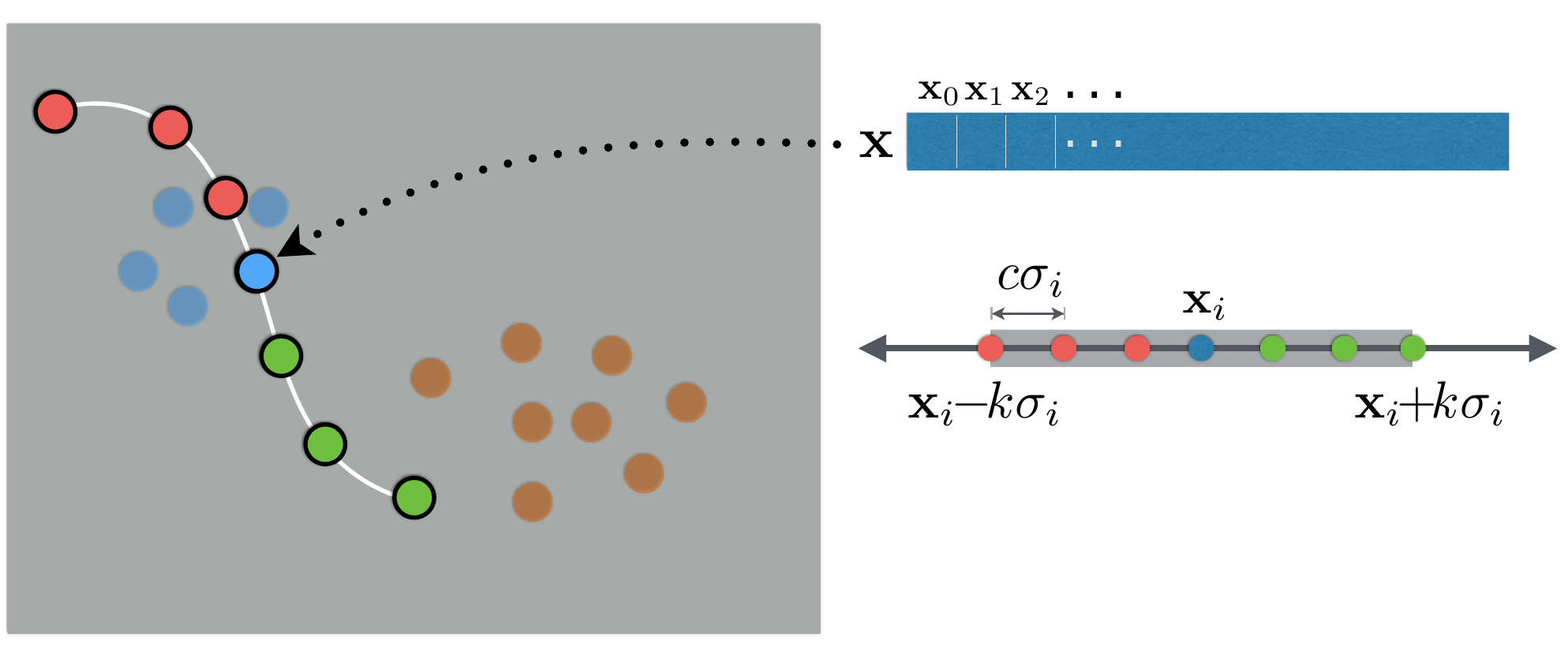}
\caption{\Pl visualize paths of forward projections.  For a given feature  
$\mathbf{x}_i$ of a data point $\mathbf{x}$, we construct a {\em proline} 
by connecting the forward projections of 
the points regularly sampled from a range of $\mathbf{x}$ values, 
where all features are fixed but $\mathbf{x}_i$ changes 
from $\mathbf{x}_i-k\sigma_i$  to $\mathbf{x}_i+k\sigma_i$.  
$\sigma_i$ is the standard deviation of the $i$th feature in 
the dataset and $k$, $c$  are constants controlling respectively the extent 
of the range and the step size with which we iterate 
over the range.\label{fig:pl}}
\end{figure}

\subsection{Forward Projection}

\Fp enables users to interactively change the feature or dimension values  of a
data point, $\mathbf{x}$, and observe how these hypothesized changes in data
modify the current projected location, $\mathbf{y}$
(Figures~\ref{fig:fp},\ref{fig:fpinaction}). This is useful because
understanding the importance and sensitivity of features (dimensions) is a key
goal in exploratory data analysis.     

We compute forward projections using out-of-sample extension (or
extrapolation)~\cite{Maaten_2009}. Out-of-sample extension is the process of 
projecting a new data point into an existing projection (e.g., learned
manifold model) using only the properties of the projection. It is
conceptually equivalent to testing a trained machine learning model with 
data that was not part of training.  

In the case of PCA, we obtain the two-dimensional position change vector $\Delta\mathbf{y}$ 
by projecting the data change vector $\mathbf{x^\prime}$ onto the principal
components: $\Delta\mathbf{y} = \Delta \mathbf{x}\;\mathbf{E}$, where     
$\mathbf{E} = \begin{bmatrix}\mathbf{e_0} & \mathbf{e_1} \end{bmatrix}$.

\begin{figure}
  {\centering
\includegraphics[width=0.5\textwidth]{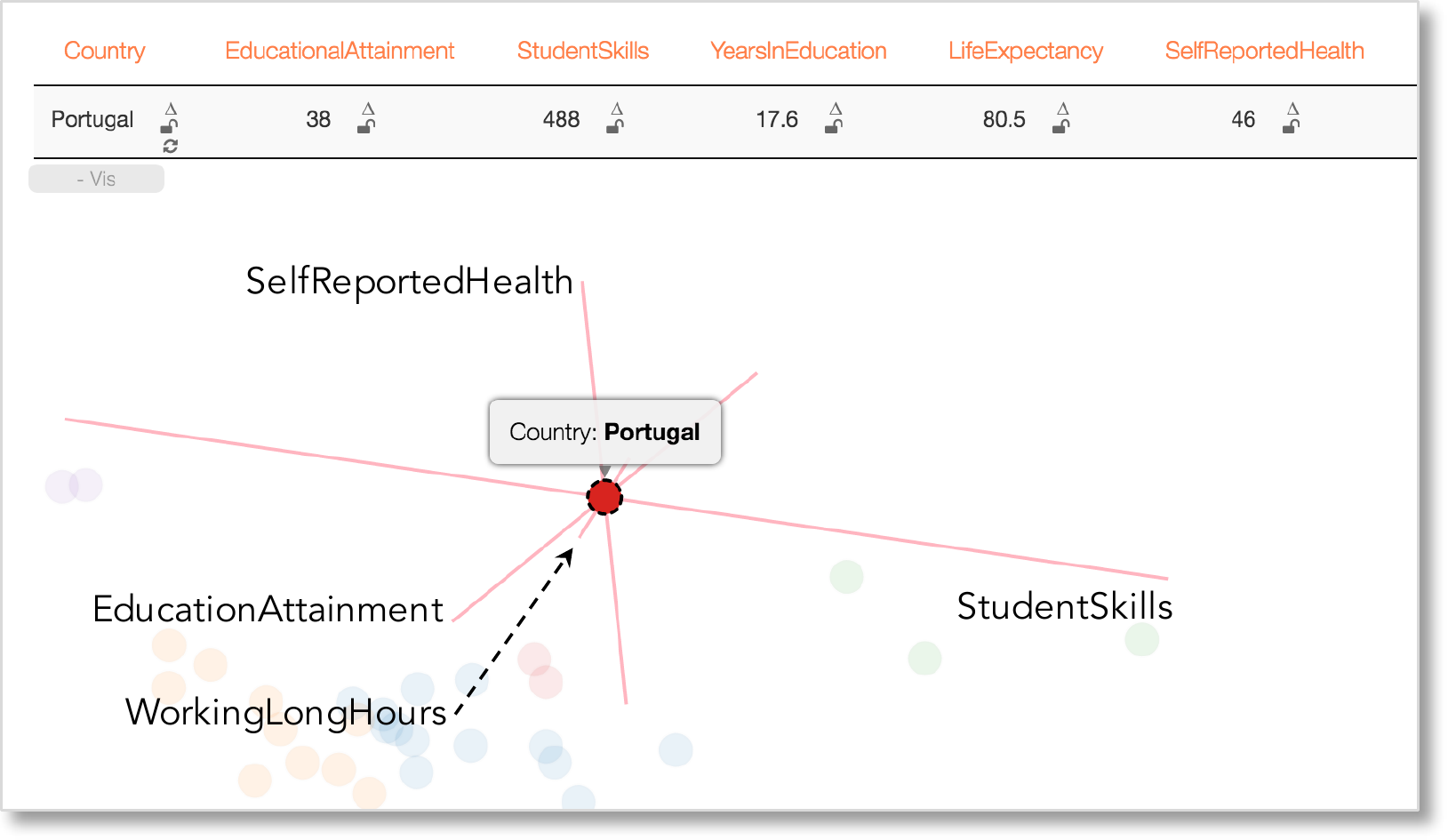}
\caption{\Pl for Portugal in a PCA projection of OECD member countries 
  based on their values for a set of development indices. 
  \Pl will be straight lines for linear dimensionality-reduction methods.
  In addition, the length of each path corresponds to the 
  speed (sensitivity, variability) along the corresponding dimension. 
   For example, {\modern StudentSkills} is the most sensitive feature 
  determining the projection in this case. Note that \fp animates 
  the speed of the change along \pl, giving the user an additional 
  cue about the importance of the dimension in the projection.  
  \label{fig:plinaction}}
}
\end{figure}

\subsection{Prolines: Visualizing Forward Projections }

It is desirable to see in advance what forward projection paths 
look like for each feature. Users can then start inspecting 
the dimensions that look interesting or important. 

\Pl visualize forward projection paths based on a range of possible values for
each feature and data point (Figures~\ref{fig:pl},~\ref{fig:plinaction}). Let
$\mathbf{x}_i$ be the value of the $i$th feature for the data point
$\mathbf{x}$. We first compute the standard deviation $\sigma_i$  for the
feature in the dataset  and devise a range $ I = \begin{bmatrix}\mathbf{x}_i -
  k\sigma_i, & \mathbf{x}_i + k\sigma_i\end{bmatrix}.$  We then iterate over
the range  with a step size of $c\sigma_i$, compute the forward projections as
discussed above, and then connect them as a path. The constants $k$, $c$
control respectively the extent of the range and the step size with which we
iterate over the range.

\Pl will be straight lines for linear dimensionality-reduction methods (Figure~\ref{fig:plinaction}), 
and therefore computing forward projections only for the extremum values of the
range $I$ is sufficient. Also note that in the case of PCA projections \pl reduces to plotting the
contributions of the feature to the principal components (loadings) as a line
vector. 

\begin{figure}[t]
\centering
\includegraphics[width=0.5\textwidth]{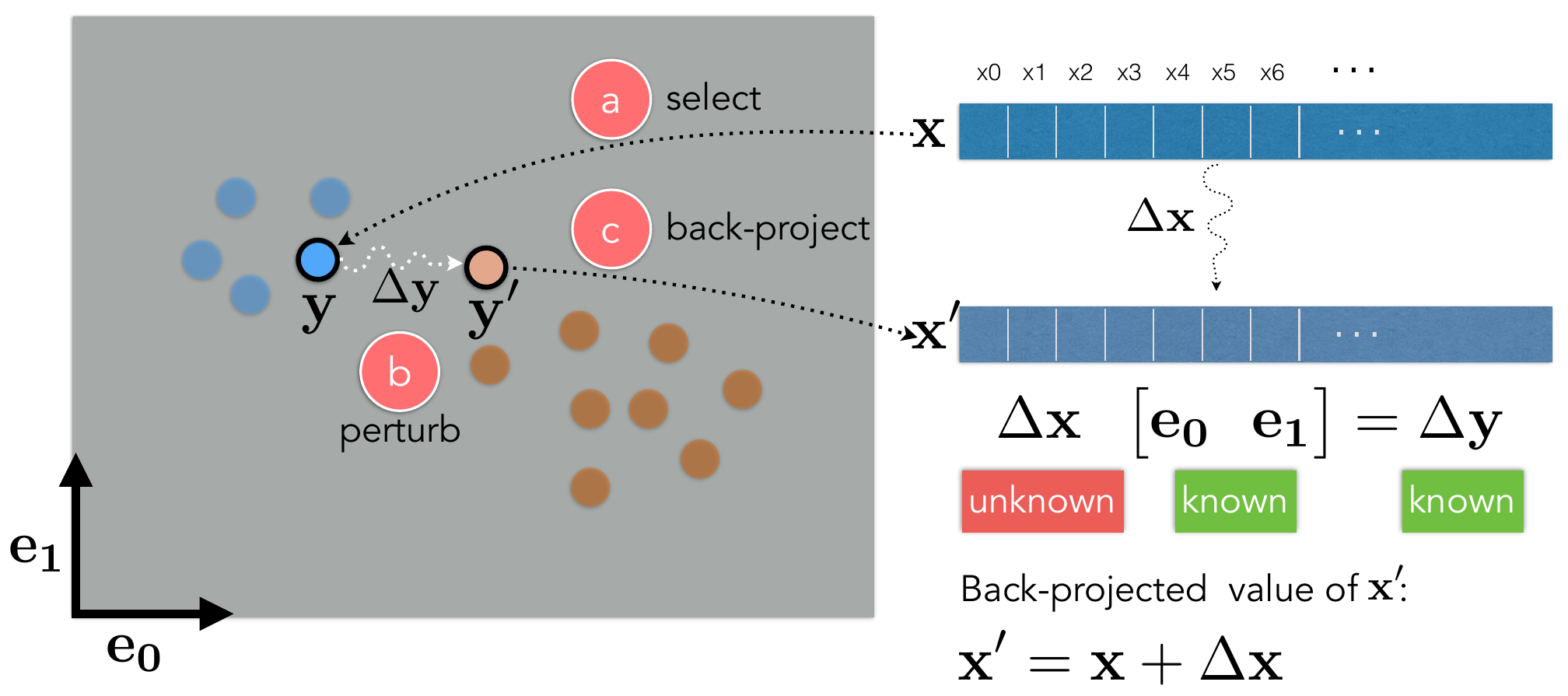}
\caption{Through \bp, users can a) select a node in the 
  projection that corresponds to a data point $\mathbf{x}$, 
  b) directly move the node in any direction 
  and c) dynamically observe what data changes $\Delta \mathbf{x}$ 
  would satisfy the hypothesized change $\Delta \mathbf{y}$ 
  in the projected position. In PCA projections, 
  $ \Delta \mathbf{x} $ can be obtained by solving for it in the linear 
  equation $\Delta\mathbf{x}\left[\mathbf{e_0}\;\;\mathbf{e_1}\right] 
  = \Delta \mathbf{y}$. \label{fig:bp}}
\end{figure}
\begin{figure}[t]
\centering
\includegraphics[width=0.5\textwidth]{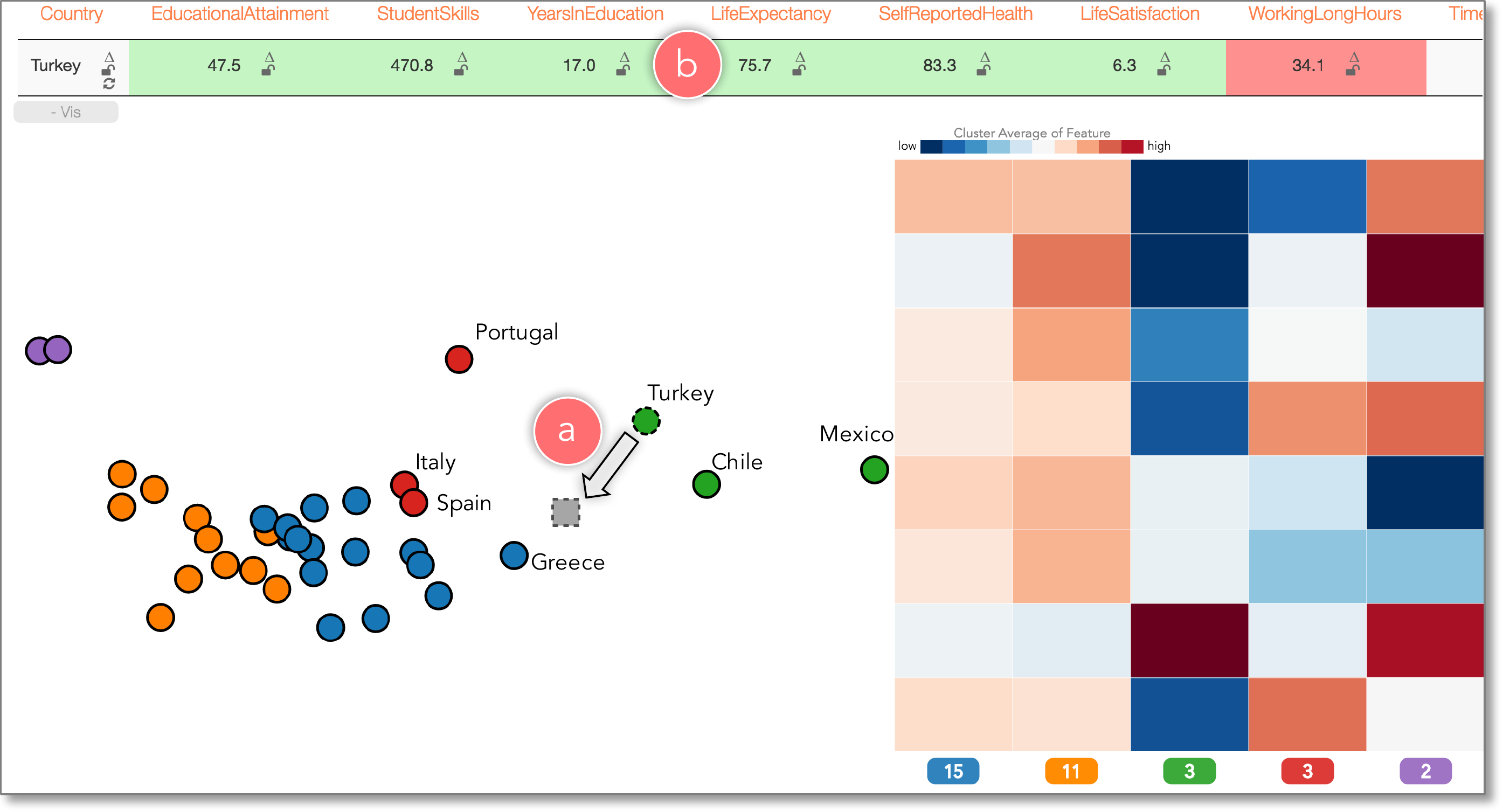}
\caption{Unconstrained \bp. A user, curious about the projection 
  difference between Turkey and Greece, a) moves the proxy node 
  for Turkey (gray square with dashed border) towards Greece.  
  The feature values for Turkey are automatically updated to satisfy 
  the new projected position as the node is moved. 
  The user b) observes that, as Turkey gets closer to 
  Greece, {\modern WorkingLongHours} 
  decreases (encoded with red) while {\modern EducationAttainment, StudentSkills, 
  YearsInEducation, LifeExpectancy, SelfReportedHealth,} and  {\modern LifeSatisfaction} 
    increase (green). {\modern TimeDevotedToLeisure} (not seen)  
    stays constant (gray). \label{fig:bpinaction}}
\end{figure}
\subsection{Backward Projection}

\begin{figure}
  \centering 
  \includegraphics[width=0.5\textwidth]{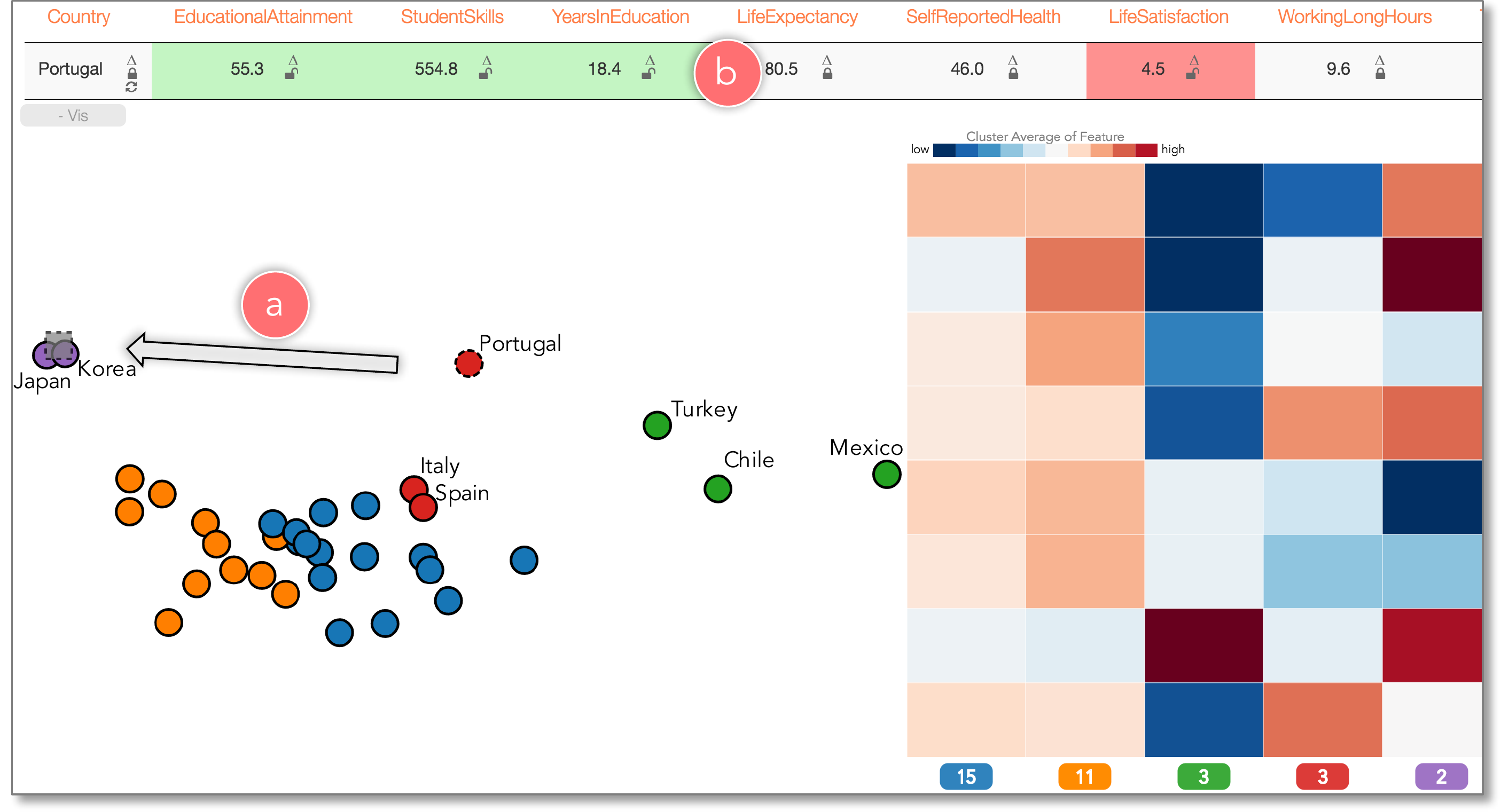}
  \caption{Constrained \bp. A user explores the projection 
  difference between Portugal and Korea, first fixing (i.e., setting equality 
  constraints) all dimensions but {\modern EducationAttainment, StudentSkills, 
    YearsInEducation, LifeSatisfaction} and then a) moving the proxy node for 
    Portugal nearer to Korea. The user b) observes that {\modern LifeSatisfaction} 
    decreases while {\modern EducationAttainment, StudentSkills,} 
    and {\modern YearsInEducation} increase. \label{fig:bpinaction-constrained}}
\end{figure}

\Bp as an interaction technique is a natural complement of \fp. Consider the
following scenario: a user looks at a projection and, seeing a cluster of points and
a single point projected far from this group, asks what changes in the
feature values of the outlier point would bring the apparent outlier near the
cluster.  Now, the user can play with  different dimensions using \fp to move the
current projection of the outlier point near the cluster. It would be 
more natural, however, to move the point directly and observe the change
(Figures~\ref{fig:bp}, \ref{fig:bpinaction}, \ref{fig:bpinaction-constrained}). 

The formulation of \bp is the same as that 
of \fp:  $\Delta \mathbf{y} = \Delta \mathbf{x}\;\mathbf{E}$.
In this case, however, $\Delta \mathbf{x}$ is unknown and we 
need to solve the equation.  

\begin{figure}
\centering
\includegraphics[width=0.5\textwidth]{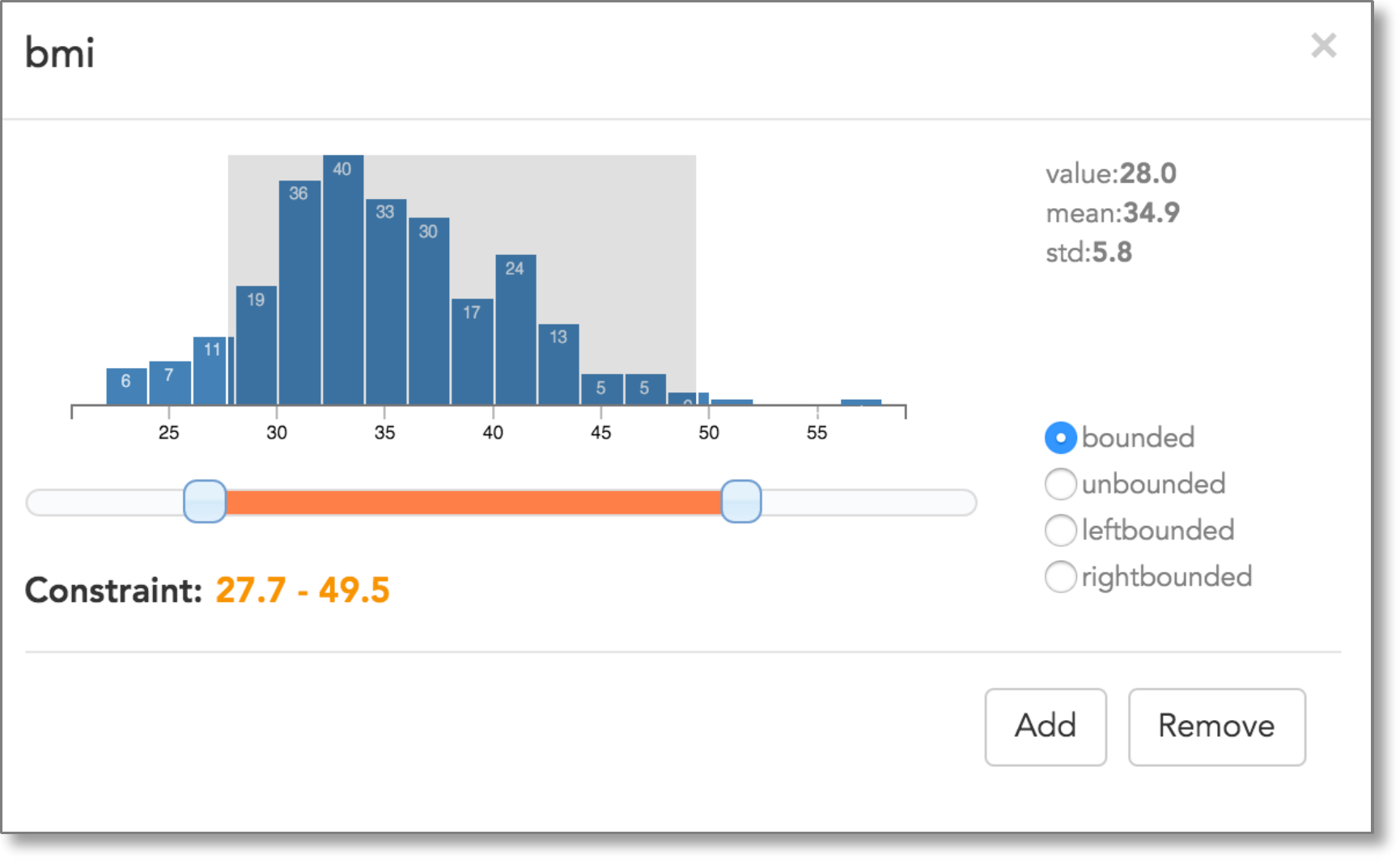}
\caption{
  Clustrophile interface for entering inequality constraints for \bp. Users can enter 
  bounded, left  and right bounded  interval constraints. The histogram shows the 
  distribution of the future ({\modern bmi}, body mass index, in this case) for 
  which the constraints are entered. The user can adjust the constraints interactively 
  using the histogram brush or the slider. 
  \label{fig:constdialog}}
\end{figure}

As formulated, the problem is underdetermined and, in general, there can be
infinitely many data points (feature values) that project to the same 
planar position. Therefore, our implementation in Clustrophile supports  both 
unconstrained and constrained backward projections. Users can introduce 
equality as well as inequality constraints (Figure~\ref{fig:constdialog}).  

In the case of unconstrained backward projection, we find $\Delta \mathbf{x}$ 
by solving a regularized least-squares optimization problem. 
\begin{equation*}
  \begin{aligned}
    & \underset{\Delta\mathbf{x}}{\text{minimize}}
    & & \|{\Delta\mathbf{x}}\|^2 \\
    & \text{subject to}
    & & \Delta\mathbf{x}\;\mathbf{E} = \Delta\mathbf{y}
  \end{aligned}
\end{equation*}
Note that this 
is equivalent to setting $\Delta \mathbf{x}= \Delta\mathbf{y}\;\mathbf{E}^T$. 
In general, for linear projections we have the unconstrained back projection 
directly.
  
As for constrained backward projection, we find $\Delta \mathbf{x}$ 
by solving the following quadratic optimization problem:   
\begin{equation*}
  \begin{aligned}
    & \underset{ \Delta\mathbf{x}}{\text{minimize}} & & \|{\Delta\mathbf{x}}\;\mathbf{E} - \Delta\mathbf{y}\|^2 \\
    & \text{subject to} & & \mathbf{C}\Delta\mathbf{x}  = \mathbf{d}\\
    & & & \mathbf{lb} \leq \Delta\mathbf{x} \leq \mathbf{ub} 
  \end{aligned}
\end{equation*}
$\mathbf{C}$ is the design matrix of equality constraints, $\mathbf{d}$ is the 
constant vector of equalities,  and $\mathbf{lb}$ and $\mathbf{ub}$ are the vectors 
of lower and upper boundary constraints.      


\subsection{System Details} 

\begin{figure}[t]
\centering
\includegraphics[width=0.5\textwidth]{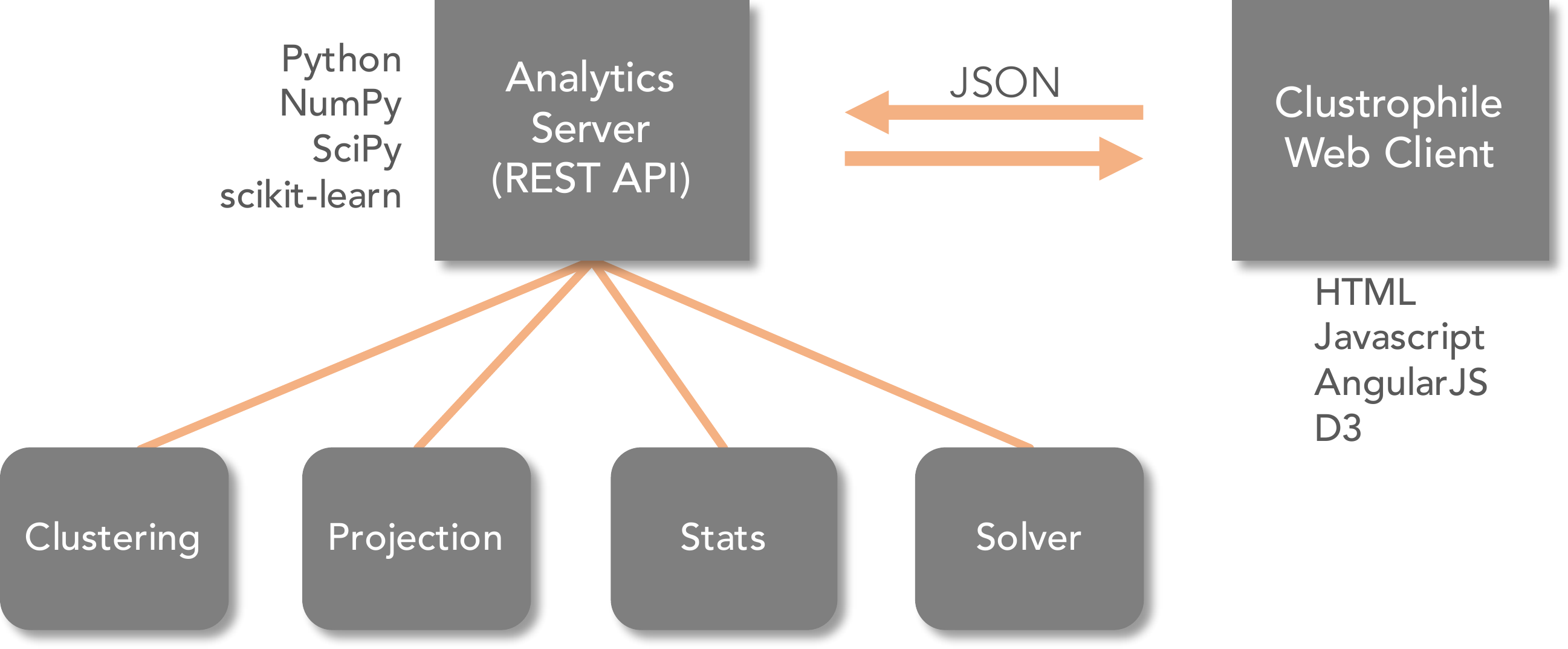}
\caption{Clustrophile architecture.\label{fig:system}}
\end{figure}

Clustrophile is a web application based on a client-server model
(Figure~\ref{fig:system}). We implemented Clustrophile's web interface
in Javascript with help of D3~\cite{d3} and 
AngularJS~\cite{angular} libraries. We generated the parser 
for the mini-language used to filter data with PEG.js~\cite{peg}.   
Most of the analytical computations are performed on Clustrophile's
Python-based analytics server, which has four modules:
clustering, projection, statistics, and solver. These modules are mainly
wrappers, making heavy use of SciPy~\cite{scipy}, NumPy~\cite{numpy}, 
and scikit-learn~\cite{scikit-learn} Python libraries. The solver 
module uses CVXOPT~\cite{cvxopt} for quadratic programming.

\section{User Feedback} 
Clustrophile is a research prototype under development and has been used over
several months by data scientists and researchers in the healthcare
domain. While we have not conducted a formal study, we briefly discuss the
informal feedback we have gathered.

Users cared most about the time-saving aspects of Clustrophile.  They were
pleased with the ability to explore different clustering and projection
algorithms and parameters without going back to their scripts.  Similarly,
among Clustrophile's favorite functionalities were the ability to add and
remove features and iteratively recompute clusterings and projections on
filtered data while staying in the context of data analysis session.  We found
that our users were more familiar with clustering than projection; indeed, for
some the relation between clustering and projection view was not always clear.

The most important request from our users was scalability. As soon as they
started using Clustrophile in commercial projects, they realized that they
wanted to be able to analyze large datasets without losing Clustrophile's current
interactive and iterative user experience (more on this in the
following section). 

\section{Sampling for Scale} 
The power of visual analysis tools such as Clustrophile comes from both
facilitating iterative, interactive analysis and leveraging visual 
perception. Exploring large datasets at interactive rates, which 
typically involves coordination of multiple visualizations through 
brushing and linking and dynamic filtering, is, however, a challenging problem. 
One source of the challenge is the cost of interactive computation
and rendering. Another is the perceptual and cognitive cost (e.g., clutter)
users incur when dealing with large numbers of visual elements.

There are two basic approaches to this problem: precomputation and
sampling~\cite{Hellerstein_2015}. Precomputation involves
processing data into a form (typically tiles or cubes) to 
interactively answer queries (e.g., zooming, panning, brushing, etc.) 
that are known in advance.  This approach has been the prevalent 
method both in the visualization community and the database community, 
from which most of the current techniques originate from.
However, precomputation is not always feasible or, indeed, desirable.  Scalable
visualization tools based on precomputation are typically applied to the
visualization of low-dimensional, spatial (e.g., map) datasets as
precomputation is infeasible when the data is high dimensional, quickly
expanding the combinatorial space of possible cubes or tiles. And, in general,
precomputation is inflexible as it restricts the ability to run arbitrary
queries. 

Sampling, considering only a selected subset of the data at a time for
analysis, is an attractive alternative to precomputation for scaling
interactive visual analytics tools.  Sampling has generality and the advantage
of easing computational and perceptual/cognitive problems at once. 
In principle, there is no reason that sampling-based visual analysis should not
be a viable and practical option. In the end, the field of statistics builds on
the premise that one can infer properties of a population (read complete data)
from its samples.  There are, however, two major challenges that, we believe, 
also limit wider adoption of sampling in general~\cite{Hellerstein_2015}. 

First is a concern about potential biases introduced by sampling. This concern
seems, however, to be at least partly unfounded, since neither aggregation bias
of precomputation nor sampling bias of complete data appear to cause as much
concern. In recent work, Kim \etal improve the effectiveness (and
trustworthiness) of sampling-based visualizations by guaranteeing the
preservation of relations (e.g., ranking) within the complete
data~\cite{Kim_2015}. The second challenge is the lack of understanding how
users interact with sampling in visual analytics tools or how sampling affects
the user experience and comprehension. Can we develop models of user behavior
regarding sampling?  How can we improve the user experience with sampling
through visualization and interaction? How can users control the sampling
process without being experts in statistics? 

Addressing these challenges would accelerate the adoption of sampling and improve 
the utilization of the unique opportunity that sampling provides in enabling visual
analysis on large datasets without losing the power of iterative, interactive
visual-analysis workflow that tools like Clustrophile facilitate. 

\section{Reflections on Projections} 

Using out-of-sample extrapolation, \fp avoids re-running
dimensionality-reduction algorithms. From the visualization point of view, this
is not just a computational convenience but also has perceptual and cognitive
advantages such as preserving the constancy of scatter-plot representations.
For example, re-running (training)  a dimensionality-reduction algorithm with
addition of a new sample can significantly alter a two-dimensional scatter plot
of the dimensionally-reduced data, despite all the original inter-datapoint
similarities stay unchanged. Many of the dimensionality-reduction algorithms
are based on eigenvector computations. Even different runs on the same dataset
can result in different---typically, flipped---planar coordinates (if
$\mathbf{v}$ is an eigenvector of a matrix so is $-\mathbf{v}$ ).   

What about interacting with nonlinear dimensionality reductions?  There are
out-of-sample extrapolation methods for many nonlinear dimensionality-reduction
techniques that make the extension of \fp with \pl  possible~\cite{Bengio_2004}.
As for \bp,  its computation will be direct in certain cases (e.g., when an
autoencoder is used).  In general, however, some form of constrained optimization
specific to the dimensionality-reduction algorithm will be needed.
Nonetheless, it is highly desirable to develop general methods that apply 
across dimensionality-reduction methods. 

\section{Visual Analysis Is Like Doing Experiments}

Data analysis is an iterative process in which analysts essentially run mental
experiments on data, asking questions and (re)forming and testing hypotheses.
Tukey and Wilk~\cite{Tukey_1965} were among the first to observe the
similarities between data analysis and doing experiments. They list eleven
similarities, for example, ``Interaction, feedback, trial and error are
all essential; convenience dramatically helpful.'' 
Albeit often implicitly, the visualization literature makes a strong case 
for designing visual analysis tools to support quick, iterative analysis 
flow that is conducive to hypothesis generation and 
testing (e.g., ~\cite{Kandel_2012}).  

We integrate our spatial interaction techniques for exploring and reasoning with 
dimensionality reductions into Clustrophile, which uses familiar data-mining and 
visualization methods  to facilitate iterative, interactive clustering analysis. 
Injecting new techniques into familiar workflows is an effective way for assessing 
their usefulness and adoption. Tukey and Wilk make an important observation on the
adoption of new techniques as part of their analogy: ``There can be great gains
from adding sophistication and ingenuity \ldots to our kit of tools, just as
long as simpler and more obvious approaches are not neglected.'' 

It is a standard practice to design visualization tools by considering criteria
determined to support user tasks. While this approach is necessary for creating
useful tools, our experience in developing Clustrophile suggests that the
design process can benefit from the regulating clarity of general, higher-level
conceptual models. To explore and reason about data, analysts generally have the
basic data-mining and visualization techniques.  They often, however, lack
interactive tools integrating these techniques to facilitate quick, iterative
what-if  analysis. Extending Tukey and Wilk's analogy between data analysis and running
experiments to visual analysis, {\em visual analysis like doing experiments},
provides a useful conceptual model for a large segment of visual analytics 
applications. Clustrophile, along with \fp, \bp and \pl, contributes to the
kit of tools needed to facilitate performing visual analysis in a similar way
to running experiments.    

\bibliographystyle{abbrv} 
\bibliography{paper} 
\end{document}